\documentclass[aps,prl,a4paper,twocolumn,longbibliography,superscriptaddress]{revtex4-2}
\usepackage[utf8]{inputenc}
\usepackage{amsmath,amsfonts,amssymb}
\usepackage{mathtools}
\usepackage{graphicx,xcolor}
\usepackage[space]{grffile}   
\usepackage[normalem]{ulem}     

\newcommand{\beq}{\begin{equation}}
\newcommand{\eeq}{\end{equation}}
\newcommand{\bei}{\begin{itemize}}
\newcommand{\eei}{\end{itemize}}
\newcommand{\ben}{\begin{enumerate}}
\newcommand{\een}{\end{enumerate}}

\newcommand{\be}{{\mathbf e}}

\usepackage{hyperref}
\hypersetup{
   pdfpagemode=UseNone, 
   pdfstartpage=1,
   pdfmenubar=true,
   pdftoolbar=true,
   colorlinks=true,
   linkcolor=blue,
   citecolor=blue,
   urlcolor=blue,
   bookmarksopen=false
 }

\definecolor{darkblue}{rgb}{0.,0.24,0.51}
\definecolor{britishracinggreen}{rgb}{0.0, 0.26, 0.15}
\definecolor{darkgreen}{rgb}{0,0.60,.2}

\def\be{\begin{equation}}
\def\ee{\end{equation}}

\def\rf#1{(\ref{#1})}

\begin{document}
\title{Heavy Fermi polarons in a one-dimensional harmonic trap}
\author{Nikolay Yegovtsev}
\affiliation{Department of Physics and Astronomy and IQ Initiative, University of Pittsburgh, Pittsburgh, Pennsylvania 15260, USA}

\date{\today}
	
\begin{abstract}
We provide an analytically tractable toy model of an infinitely heavy impurity interacting with the spin-polarized gas of bath fermions via a contact potential in 1D and placed in a harmonic trap. The solution to this problem requires knowledge of the single-particle fermionic eigenstates in the presence of a harmonic trap and a contact potential. We show how the spectrum can be understood from a perturbative solution of the exact transcendental equation in the weak and strong-coupling regimes. We additionally provide expressions for normalized wavefunctions and different overlaps between the states in the presence and absence of the impurity. Using these exact results, we analyze the energy of the polaron, derive Tan's contact-like relation for the density at the center of the harmonic trap, and compute the quasiparticle residue.
\end{abstract}
\maketitle



The problem of a mobile impurity introduced into a polarized Fermi gas is called the Fermi polaron problem, and it has gained a lot of attention in recent years in both theoretical and experimental communities \cite{massignan2025polarons}. Despite its simplicity, there is no exact solution to this problem except for the equal-mass case in 1D \cite{mcguire1966interacting, guan2013fermi}. On the other hand, if the mass of the impurity $M$ is taken to infinity, then the problem becomes exactly solvable in any number of dimensions with the help of Fumi's theorem \cite{mahan2013many, combescot2007normal, giraud2009highly}. Despite its simplicity and power, Fumi's theorem is limited to the study of fermions in free space; however, in many experimentally relevant scenarios, the system is placed in a trapping potential, which, to a good approximation, can be treated as harmonic. The properties of such a polaron will now depend on the single-particle energy levels in the harmonic trap plus the impurity potential. This problem is closely related to the problem of two particles in a harmonic trap interacting via contact potential~\cite{busch1998two}. Recently, \cite{koscik2018exactly, koscik2019exactly} have discussed the analytical solution for the case of two particles in the harmonic trap interacting by a soft-core potential, which is relevant for the scattering of Rydberg atoms, and, in principle, for the Rydberg polarons \cite{sous2020rydberg}. While both~\cite{koscik2018exactly} and~\cite{koscik2019exactly} presented the formal solution in terms of the exact transcendental equation that describes the spectrum of the eigenstates, the analysis was mostly done numerically. Interestingly, such transcendental equations can be approximately solved analytically in the regimes of weak and strong interactions between two particles in a systematic way. This approach gives a better understanding of the structure of the single-particle energy levels, which is required for the analytical analysis of the heavy Fermi polarons in harmonic traps. Such an analysis is also relevant to a broader class of problems in atomic physics, because the two-particle scattering in a harmonic trap is quite ubiquitous \cite{giorgini2008theory, PhysRevResearch.2.023108, 8mnc-x42q}.

In this Letter, we consider an infinitely heavy impurity interacting with a harmonically confined polarized Fermi gas in 1D, and show that it can be analyzed analytically in various regimes. First, we show how this problem can be obtained by a suitable decomposition of the many-body Hamiltonian for the mobile impurity, where the effects of finite mass can be treated perturbatively. Then we study the single-particle energy levels of the resultant system in the weak and strong-coupling regimes and derive analytical expressions for the wavefunction overlaps that are useful for the study of the dynamics of harmonically confined one-dimensional interacting gases. Finally, we use the obtained results to investigate the static quasiparticle properties of the polaron, such as its energy, contact, and quasiparticle residue, which can be measured experimentally from the absorption spectrum and the density measurement at the position of the impurity. We note the special role that the harmonic trap plays in our analysis. First, it makes all single-particle eigenstates mass-dependent. That is, in the limit $M\to \infty$, they become automatically $\delta$-function localized, which allows us to develop a finite mass expansion around the exactly soluble $M\to\infty$ point. Second, for a finite strength of the fermion-impurity interaction, it prevents the Anderson orthogonality catastrophe (OC) from happening, so the polaron quasiparticle residue remains finite as we increase the number of particles in the system. Here, OC only emerges in the limit of infinite interactions. Our analytic results provide new insight into the physics of heavy impurities interacting with bath particles in a harmonic trap.  They are relevant to the study of polarons in large mass-imbalanced atomic mixtures \cite{PhysRevLett.95.170408, PhysRevA.84.011606, PhysRevLett.106.205304, PhysRevLett.119.233401} and the impurities localized inside the species-selective optical traps \cite{ PhysRevX.12.011040, anand2024dual, PhysRevA.110.043118}, and cover both mesoscopic and macroscopic regimes of the fermionic baths. 

\textit{Reduction to the single particle problem} -- The Hamiltonian describing the system of $2N$ polarized fermions of mass $m$ interacting with a single impurity of mass $M$ via contact potential of strength $g$ is given by:
\begin{equation}
\label{eq:Hamiltonian}
\begin{split}
H &=    \sum_{i=1}^{2N}\frac{p_i^2}{2m}+\sum_{i=1}^{2N}\frac{m\omega^2x_i^2}{2}+g\sum_{i=1}^{2N}\delta(x_i-X)\\
&+\frac{P^2}{2M} +\frac{M\omega^2X^2}{2}.
\end{split}
\end{equation}
We chose to work with $2N$ particles because odd eigenstates will not contribute to the final results. Eq.~\rf{eq:Hamiltonian} can be decomposed as:
\begin{equation}
\begin{split}
H_0 = &  \frac{P^2}{2M} +\frac{M\omega^2X^2}{2} \\
&+ \sum_{i=1}^{2N}\frac{p_i^2}{2m} + \sum_{i=1}^{2N}\frac{m\omega^2x_i^2}{2} + g\sum_{i=1}^{2N}\delta(x_i),\\
&H_\text{int} = g\sum_{i=1}^{2N}\left[\delta\left(x_i-X\right)-\delta(x_i)\right].
\end{split}    
\end{equation}
$H_0$ describes a single impurity inside the harmonic trap and the collection of fermions in the same trap that also experience the presence of the contact potential at the origin. The ground state of $H_0$ corresponds to the product of the ground state wavefunction of the harmonic oscillator that describes the impurity and the Slater determinant constructed out of the single-particle eigenstates of the harmonically confined fermions scattering on the contact potential. In the limit $M\to \infty$, the impurity's wavefunction becomes localized at the origin, so the expectation value of $H_\text{int}$ in this ground state is identically zero. For finite and large $M$, this expectation will depend on $m/M$ in a nontrivial way, yet by construction, this correction will be small. Such a reduction scheme can be applied in any dimension and to an arbitrary fermion-impurity potential. Here we only focus on the nontrivial part of $H_0$, which describes fermions, and leave the detailed perturbative treatment of $H_\text{int}$ for future work. Thus, the problem in Eq.~\rf{eq:Hamiltonian} reduces to finding the single-particle fermionic eigenstates in the presence of the harmonic trap and a contact potential placed at the origin. Some partial results on the properties of the energies and eigenstates appeared in previous studies \cite{bera2008perturbative,viana2011solution, akyuz2024harmonic,donelli2025impactquantumcoherencedynamics} in a different context. Here, we present old and new results in a unified fashion, showing how they can be obtained systematically. All single-particle eigenstates are either even or odd, but odd states are unaffected by the presence of a contact potential, so our focus will be on even states. The wave function is given by Tricomi's confluent hypergeometric function $U(\alpha, \gamma, x)$ \cite{abramowitz1948handbook, SuppMat}:
\begin{equation}
\label{eq:wtricomi}
\varphi(x) = A U(\alpha, \frac{1}{2},\lambda x^2)e^{-\frac{1}{2}\lambda x^2},    
\end{equation}
where $\lambda = m\omega/\hbar$, and $A$ is the normalization constant that will be computed later. Here $\alpha$ is the dimensionless parameter of the hypergeometric function that is related to the energy via \cite{flugge2012practical, SuppMat}: 
\begin{equation}
\label{eq:energy}
E = 2\hbar\omega\left(\frac{1}{4}-\alpha\right).    
\end{equation}
One can derive the exact transcendental equation that describes the spectrum in terms of gamma functions \cite{viana2011solution, SuppMat}:
\begin{equation}
\label{eq:spectrum}
\begin{split}
&2\alpha = -\kappa\frac{\Gamma(\alpha+1)}{\Gamma(\alpha+\frac{1}{2})},  \\ 
&\kappa = \sqrt{\frac{m}{\hbar^3\omega}}g.
\end{split}
\end{equation}
In general, we need to solve the above equation numerically, as is typically done; however, we can still understand the behavior of the spectrum by studying this equation for $\kappa\ll1$, and $\kappa\gg1$ by perturbatively solving Eq.~\rf{eq:spectrum} and using the property of the gamma function that it does not have zeros on the real axis and its poles are located at negative integers and zero \cite{whittaker2020course}.

\textit{Weak coupling expansion} -- In the absence of the interactions $g=0$, the spectrum is that of the even states of the harmonic oscillator, so $\alpha = -n$ \cite{flugge2012practical, SuppMat}. For $\kappa\ll1$ we can seek the solution in the form $\alpha = -n+\kappa c_1+\kappa^2c_2+\cdots$ to the second order in $\kappa$, where $c_1, c_2$ are coefficients that can be determined when expanding both sides of Eq.~\rf{eq:spectrum} to first order in $\kappa$. After some algebra, we arrive at the following expression for the energy of the eigenstate \cite{bera2008perturbative, SuppMat}:
\begin{equation}
\label{eq:weakcoupling}
\begin{split}
\frac{E_n}{\hbar\omega} = & 2n+\frac{1}{2} +\frac{\kappa}{\pi}\frac{\Gamma(n+\frac{1}{2})}{\Gamma(n+1)}\\
&- \frac{\kappa^2}{2\pi^2}\left[\frac{\Gamma(n+\frac{1}{2})}{\Gamma(n+1)}\right]^2\left(\psi^{(0)}(n+1)-\psi^{(0)}(n+\frac{1}{2})\right).    
\end{split}
\end{equation}
where $\psi^{(0)}(x) = \Gamma'(x)/\Gamma(x)$ is the digamma function. The second-order term in $\kappa$ gives the final result of summation over intermediate states when one attempts to solve the problem using the standard QM second-order perturbation theory \cite{bera2008perturbative}. In principle, such a method can also be applied to similar problems~\cite{koscik2018exactly, koscik2019exactly}, where the exact transcendental equation for the spectrum is known. Going beyond the second order can be done in a straightforward way. From the perturbative solution of the transcendental equation, we can see that for the generic state ($n\neq 0$), the applicability of the first-order perturbation theory is that $|c_1\kappa|\ll n$, which implies $\kappa \ll 2\pi n \Gamma(n+1)/\Gamma(n+\frac{1}{2})\approx2\pi n^{3/2}$, so, for sufficiently large $n$ the first-order perturbation theory works well even for $\kappa$ that is not necessarily small. This is analogous to the applicability of the Born approximation to the scattering of two particles with large incoming momenta, even when the interaction potential itself may not be weak \cite{landau2013quantum}. The corresponding criterion for the ground state ($n=0$) is the smallness of the second-order result compared to the first-order one: $|c_1\kappa|\ll |c_2\kappa^2|$, which gives $\kappa\ll 2.56$.

\begin{figure}[t]
    \centering
    \includegraphics[width=\columnwidth]{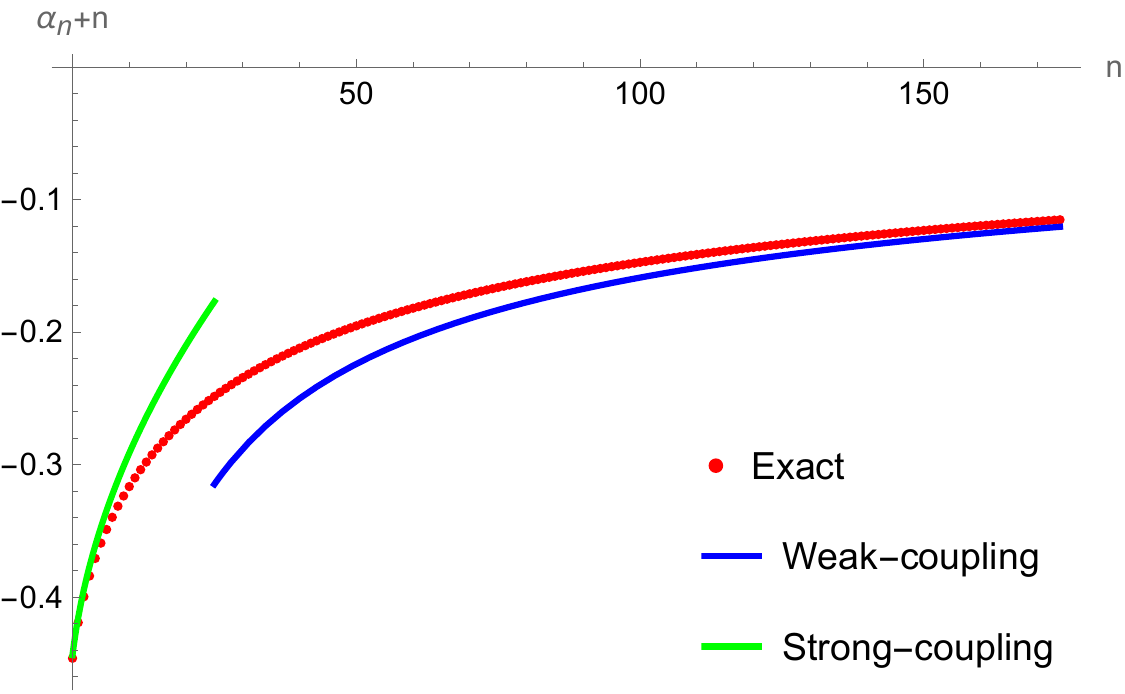}
    \caption{$\alpha_n+n$ for the case $\kappa=10$. It is extracted from Eq.~\rf{eq:energy} and the expressions for the weak and strong-coupling regimes are given by Eq.~\rf{eq:weakcoupling} and Eq.~\rf{eq:strongcouplingr}.}
    \label{fig:1}
\end{figure}

\textit{Strong coupling expansion} -- When $\kappa \to -\infty$, one option is to have $\alpha$ large and positive, and another is to be close to $-n+\frac{1}{2}$ to compensate for the term on the right side in Eq.~\rf{eq:spectrum}. Let us first discuss the case of $\alpha\gg1$. In this scenario, the ratio of two gamma functions goes as $\sqrt{\alpha}$, so the leading-order result is $\alpha = \kappa^2/4$, which corresponds to the energy of the ground state of the pure delta potential.  The expansion goes in inverse powers of $\kappa^2$ with the leading-order result in $1/\kappa^2$:
\begin{equation}
\label{eq:boundstate}
\frac{E}{\hbar\omega} = -\frac{\kappa^2}{2} +\frac{1}{4\kappa^2}+\cdots    
\end{equation}
This result is quite robust, as even for $\kappa = -2$ it gives a relative error smaller than $0.2\%$. The analysis of the negative half-integer scenario is the same for both attractive and repulsive interactions $\alpha = -n\pm\frac{1}{2}+\frac{c_1}{\kappa} + \frac{c_2}{\kappa^2}+\cdots$, where plus corresponds to the attractive and minus to the repulsive interactions. For the repulsive case \cite{SuppMat}:
\begin{equation}
\label{eq:strongcouplingr}
\begin{split}
\frac{E_n}{\hbar\omega} &= 2n+\frac{3}{2} -\frac{4}{\kappa\pi}\frac{\Gamma(n+\frac{3}{2})}{\Gamma(n+1)}\\
& -\frac{8}{\kappa^2\pi^2}\left[\frac{\Gamma(n+\frac{3}{2})}{ \Gamma(n+1)}\right]^2\left(\psi^{(0)}(n+1)-\psi^{(0)}\left(n+\frac{3}{2}\right)\right).
\end{split}
\end{equation}
For the attractive case \cite{SuppMat}:
\begin{equation}
\label{eq:strongcouplinga}
\begin{split}
\frac{E_n}{\hbar \omega} &=2n-\frac{1}{2} - \frac{4}{\kappa\pi}\frac{\Gamma(n+\frac{1}{2})}{\Gamma(n)}\\
&-\frac{8}{\kappa^2\pi^2}\left[\frac{\Gamma(n+\frac{1}{2})}{ \Gamma(n)}\right]^2\left(\psi^{(0)}(n)-\psi^{(0)}\left(n+\frac{1}{2}\right)\right).
\end{split}
\end{equation}
We observe that the strong-coupling expansion works well for low-lying states, and gets progressively worse for large $n$, where one can use the weak-coupling results instead. This behavior is shown in Fig.~\ref{fig:1} for the case $\kappa=10$.

\textit{Wave functions overlaps} -- Normalization factor $A$ can also be computed analytically \cite{akyuz2024harmonic, SuppMat}:
\begin{equation}
\label{eq:An}
A_n = \left[\left(\frac{\lambda^{\frac{1}{2}}}{\pi}\right)\frac{\Gamma(\alpha_n)\Gamma(\alpha_n+\frac{1}{2})}{\psi^{(0)}(\alpha_n+\frac{1}{2})-\psi^{(0)}(\alpha_n)}\right]^{1/2}.    
\end{equation}
Using this, one can further compute the overlaps between the even normalized eigenstates in the presence of the impurity $|\varphi_n\rangle$ with the even eigenstates of the simple harmonic oscillator $|\phi_m\rangle$ \cite{SuppMat}:
\begin{equation}
\label{eq:hooverlap}
\langle \varphi_n|\phi_m\rangle = \frac{(-1)^m}{\alpha_n+m}\frac{A_n}{\Gamma(\alpha_n)}\left[\left(\frac{\pi}{\lambda}\right)^{\frac{1}{2}}\frac{(2m)!}{2^{2m}(m!)^2)}\right]^{\frac{1}{2}}.
\end{equation}
Such overlaps can be used to study the non-equilibrium properties of the interacting gases in the harmonic trap.

\textit{Static quasiparticle properties of the Fermi polaron} -- The energy of the polaron is given by the difference between the energy of the system in the presence of the impurity and the energy of the first $2N$ states of the harmonic oscillator. Since the odd sector does not contribute, so the energy is given by:
\begin{equation}
\label{eq:hppolener}
E  = -2\hbar\omega\sum_{n=0}^{N-1}\left(\alpha_n+n\right).
\end{equation}
While traditionally, one fixes the number of particles in the trap and varies the interaction, here we fix the interaction strength and change the number of particles in the system, which allows us to understand this model both in the mesoscopic and macroscopic regimes. 

If $|\kappa|\ll1$, then the perturbative description works for all eigenstates. $\kappa^0$ terms in Eq.~\rf{eq:hppolener} cancel out exactly, and the energy of the polaron is given by the sum of energies given in Eq.~\rf{eq:weakcoupling}, containing only terms in powers of $\kappa$. This result is valid for an arbitrary number of fermions in the system.

When $\kappa\gg1$, the energy of the polaron will strongly depend on how many particles there are in the system as can be seen from Fig.~\ref{fig:1}. Let us first consider the case of a mesoscopic system, where the number of particles is small enough for the strong coupling expansion to work uniformly for all $n\leq N-1$. Experimentally, such a scenario can arise if we load a finite number of atoms inside a small optical trap such as a tweezer. Here we can use the results of Eq.~\rf{eq:strongcouplingr} and Eq.~\rf{eq:strongcouplinga}. For the case of infinite repulsion, the fermionic wavefunctions will become zero at the origin, and the energy levels will be as of odd states of the harmonic oscillator, so the leading contribution to the energy of the repulsive polaron in this regime will be 
\begin{equation}
E = N\hbar\omega,    
\end{equation}
which can be thought of as the energy of an extra fermion placed at the lowest available energy level. $1/\kappa$ corrections will come from summing corresponding $1/\kappa$ terms in Eq.~\rf{eq:strongcouplingr}. In the case of infinite attraction, the main contribution will come from the deep bound state in Eq.~\rf{eq:boundstate}, and the difference between the terms at $\kappa^0$ order in Eq.~\rf{eq:strongcouplinga} and Eq.~\rf{eq:hppolener}:
\begin{equation}
E = -\frac{mg^2}{2\hbar^2} - (N-1)\hbar\omega,    
\end{equation}
which can be interpreted as the energy of the dimer and the hole created in the Fermi sea by dimerizing the impurity and one of the fermions \cite{8mnc-x42q}. Corrections to the energy in powers of $1/\kappa$ can be added by summing corresponding terms in Eq.~\rf{eq:strongcouplinga} and adding the $1/\kappa^2$ correction from Eq.~\rf{eq:boundstate}. Note that for the attractive case, the deep bound state is counted separately from the $n=0$ case, so the upper limit on the sum is $N-2$ instead of $N-1$. This shift is also responsible for the counting that gives the $\hbar\omega(N-1)$ term in the expression, and its physical interpretation as a hole. 

\begin{figure}[t]
    \centering
    \includegraphics[width=\columnwidth]{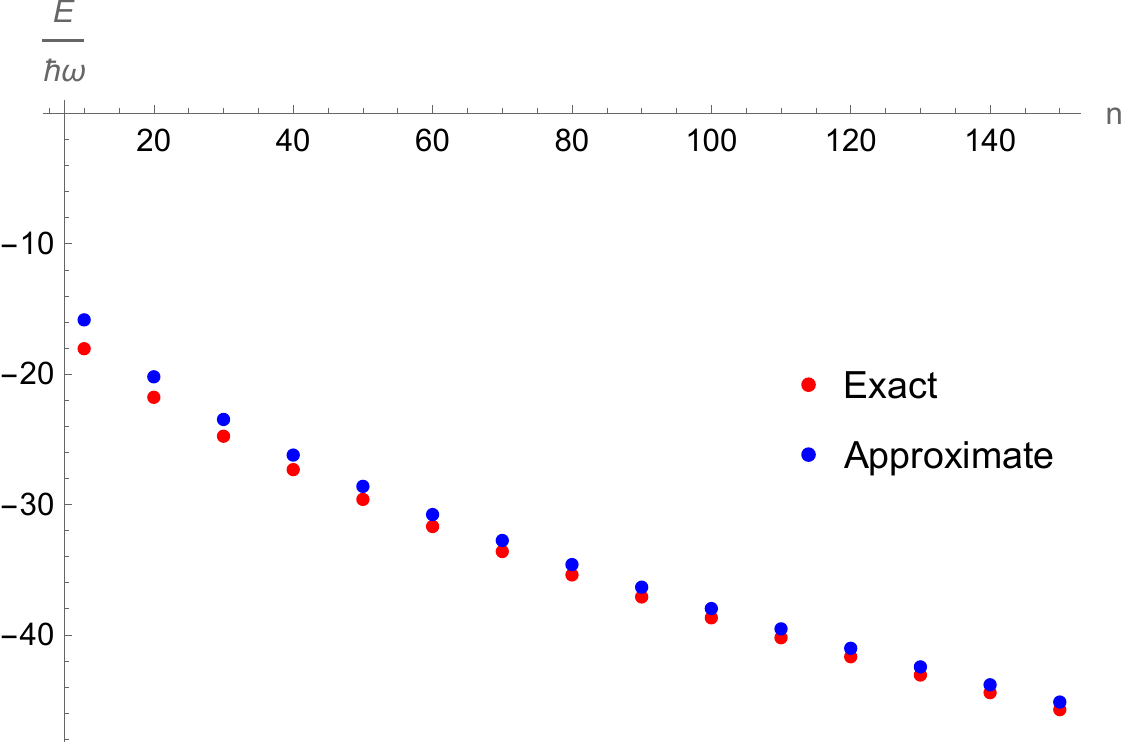}
    \caption{Energy of the attractive polaron $\frac{E}{\hbar \omega}$ for the case $\kappa=-5$. Approximate energy is comprised of the sum of the weak coupling result Eq.~\rf{eq:weakcoupling} with the lower limit on the sum set to $n_l=5$ and the energy of the bound state Eq.~\rf{eq:boundstate}.}
    \label{fig:2}
\end{figure}

In a more general case, one needs to find the energy levels numerically, so the energy of the polaron is not given by any simple expansion. We should note, however, that even if $|\kappa|\geq1$, provided that we have a lot of particles in the system, the contribution to the energy will mostly come from high $n$ states, so the expression for the polaron energy will resemble weak-coupling result. Indeed, for large values of $n$, we have:
$\Gamma(n+\frac{1}{2})/\Gamma(n+1)\sim1/ \sqrt{n}$, and by approximating sum by the integral we have:
$\sum_{n=0}^{N-1}\Gamma(n+\frac{1}{2})/\Gamma(n+1)\sim 2\sqrt{N}$. 
By comparing the energy contribution from the delta-function bound state Eq.~\rf{eq:boundstate} and the contribution from the rest of the spectrum, the contribution of the former is invisible on top of the latter provided that $\kappa\ll\sqrt{N}$.
If that is not the case, but $N\gg1$, the energy of the polaron can be approximated as the sum of the weak-coupling result in Eq.~\rf{eq:weakcoupling} and the energy of the bound state Eq.~\rf{eq:boundstate}. Since the weak-coupling perturbation theory fails for low-lying states when $\kappa\gg1$, then to fix the lower limit in the sum we can use the criterion for the applicability of the first-order perturbation theory. The Fig.~\ref{fig:2} shows how this approach works in the case $\kappa = -5$ and when $N\leq 150$. Choosing $n_l=5$ as the lower cutoff on the sum is in good agreement with the exact numerical result. 
Similar analysis holds for repulsive polarons, however, since there is no bound state of the delta potential, the result will be given just by the weak coupling perturbation theory.

We can define Tan's contact-like quantity as the derivative of the polaron energy $E$ with respect to the strength of the interaction $g$ \cite{barth2011tan}. Using Eq.~\rf{eq:hppolener}, this can be done directly at the level of the transcendental equation Eq.~\rf{eq:spectrum}. Meanwhile, the density of the fermions at the origin is given by $\rho(0) =\sum_{n=0}^{N-1}A_n^2U^2(\alpha_n,\frac{1}{2},0)$, so using the expression for the normalization factor Eq.~\rf{eq:An} and known behavior of $U(\alpha,\gamma,x)$ at the origin, we can directly show:
\begin{equation}
\label{eq:contact}
\frac{\partial E}{\partial g}  = \rho(0).   
\end{equation}
This relation is nonperturbative and can be measured experimentally by probing the density of the Fermi gas at the location of the impurity. 

\begin{figure}[t]
    \centering
    \includegraphics[width=\columnwidth]{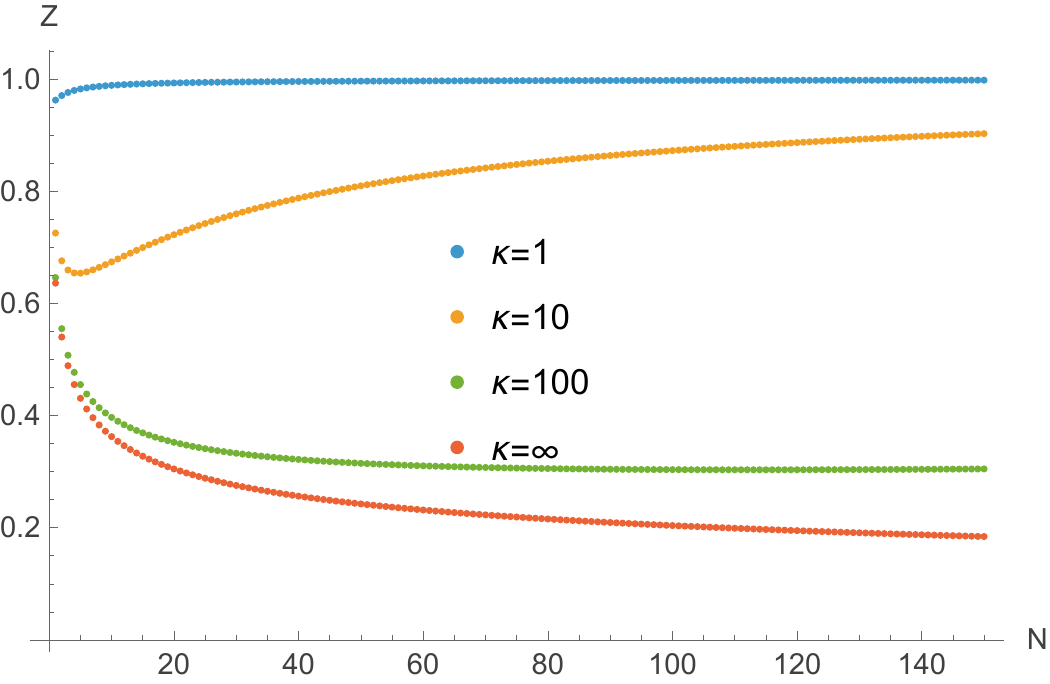}
    \caption{Quasiparticle residue $Z$ in Eq.~\rf{eq:residue} of the repulsive polaron as function of $N$ for various values of coupling strength $\kappa$.}
    \label{fig:3}
\end{figure}

Polaron's quasiparticle residue $Z$ is defined as the overlap of the ground state wavefunction of the system in the presence of the interactions $|\Phi\rangle$ with the ground state of the noninteracting model $|\Phi_\text{NI}\rangle$:
\begin{equation}
\label{eq:residue}
Z =  |\langle \Phi|\Phi_\text{NI}\rangle |^2.   
\end{equation}
This overlap reduces to the calculation of the determinant of the matrix elements of the form given in Eq.~\rf{eq:hooverlap}, so we can compute it exactly.
Fig.~\ref{fig:3} shows $Z$ as a function of $N$ for repulsive polarons for distinct values of $\kappa$, where $\kappa = \infty$ corresponds to the limit of infinite repulsion studied in \cite{levinsen2015strong} for the case of a mobile impurity. We observe that in the limit $\kappa = \infty$, $Z$ decreases with $N$ and vanishes in the $N\to\infty$ limit, signaling the Anderson orthogonality catastrophe (OC) \cite{levinsen2015strong, PhysRevLett.18.1049}. For finite and large values of $\kappa\gg1$, it initially decreases with $N$, however, it starts to increase and saturates to some finite value as we increase the number of fermions in the trap. For $\kappa=1$, $Z$ increases monotonically with $N$. The general behavior of $Z$ in our system is quite different from what happens in free space, where it decays to zero with the increase of $N$ for an arbitrary finite strength of the interaction \cite{anderson1967infrared, knap2012time, orso2025quasi}. Our results suggest that for a finite value of $\kappa$, OC will not occur, and one has to scale the strength of the fermion-impurity interaction with $N$ to see its emergence, as was noted in \cite{PhysRevLett.124.110601}. 



\textit{Conclusions and outlook} -- We analytically studied various static quasiparticle properties of an infinitely heavy impurity in a one-dimensional harmonic trap that is relevant to the study of impurities in the mass-imbalanced atomic species. The presented formalism also allows us to consider finite-range potentials and incorporate the effects of the finite impurity mass by perturbation theory around the exact solution presented here. The detailed study of these effects is left for future work.


\vspace{5mm}
\begin{acknowledgments}
We acknowledge P. Massignan, G. Astrakharchik, A. Duspayev, E. Zhao, and W. V. Liu for insightful discussions and useful comments on the manuscript. This work was supported by AFOSR Grant No.~FA9550-23-1–0598.
\end{acknowledgments}

\bibliography{UnitaryPolaron}

\begin{thebibliography}{37}%
\makeatletter
\providecommand \@ifxundefined [1]{%
 \@ifx{#1\undefined}
}%
\providecommand \@ifnum [1]{%
 \ifnum #1\expandafter \@firstoftwo
 \else \expandafter \@secondoftwo
 \fi
}%
\providecommand \@ifx [1]{%
 \ifx #1\expandafter \@firstoftwo
 \else \expandafter \@secondoftwo
 \fi
}%
\providecommand \natexlab [1]{#1}%
\providecommand \enquote  [1]{``#1''}%
\providecommand \bibnamefont  [1]{#1}%
\providecommand \bibfnamefont [1]{#1}%
\providecommand \citenamefont [1]{#1}%
\providecommand \href@noop [0]{\@secondoftwo}%
\providecommand \href [0]{\begingroup \@sanitize@url \@href}%
\providecommand \@href[1]{\@@startlink{#1}\@@href}%
\providecommand \@@href[1]{\endgroup#1\@@endlink}%
\providecommand \@sanitize@url [0]{\catcode `\\12\catcode `\$12\catcode `\&12\catcode `\#12\catcode `\^12\catcode `\_12\catcode `\%12\relax}%
\providecommand \@@startlink[1]{}%
\providecommand \@@endlink[0]{}%
\providecommand \url  [0]{\begingroup\@sanitize@url \@url }%
\providecommand \@url [1]{\endgroup\@href {#1}{\urlprefix }}%
\providecommand \urlprefix  [0]{URL }%
\providecommand \Eprint [0]{\href }%
\providecommand \doibase [0]{http://dx.doi.org/}%
\providecommand \selectlanguage [0]{\@gobble}%
\providecommand \bibinfo  [0]{\@secondoftwo}%
\providecommand \bibfield  [0]{\@secondoftwo}%
\providecommand \translation [1]{[#1]}%
\providecommand \BibitemOpen [0]{}%
\providecommand \bibitemStop [0]{}%
\providecommand \bibitemNoStop [0]{.\EOS\space}%
\providecommand \EOS [0]{\spacefactor3000\relax}%
\providecommand \BibitemShut  [1]{\csname bibitem#1\endcsname}%
\let\auto@bib@innerbib\@empty
\bibitem [{\citenamefont {Massignan}\ \emph {et~al.}(2025)\citenamefont {Massignan}, \citenamefont {Schmidt}, \citenamefont {Astrakharchik}, \citenamefont {{\.I}mamoglu}, \citenamefont {Zwierlein}, \citenamefont {Arlt},\ and\ \citenamefont {Bruun}}]{massignan2025polarons}%
  \BibitemOpen
  \bibfield  {author} {\bibinfo {author} {\bibfnamefont {P.}~\bibnamefont {Massignan}}, \bibinfo {author} {\bibfnamefont {R.}~\bibnamefont {Schmidt}}, \bibinfo {author} {\bibfnamefont {G.~E.}\ \bibnamefont {Astrakharchik}}, \bibinfo {author} {\bibfnamefont {A.}~\bibnamefont {{\.I}mamoglu}}, \bibinfo {author} {\bibfnamefont {M.}~\bibnamefont {Zwierlein}}, \bibinfo {author} {\bibfnamefont {J.~J.}\ \bibnamefont {Arlt}}, \ and\ \bibinfo {author} {\bibfnamefont {G.~M.}\ \bibnamefont {Bruun}},\ }\bibfield  {title} {\bibinfo {title} {\emph {Polarons in atomic gases and two-dimensional semiconductors}},\ }\href@noop {} {\bibfield  {journal} {\bibinfo  {journal} {arXiv preprint arXiv:2501.09618}\ } (\bibinfo {year} {2025})}\BibitemShut {NoStop}%
\bibitem [{\citenamefont {McGuire}(1966)}]{mcguire1966interacting}%
  \BibitemOpen
  \bibfield  {author} {\bibinfo {author} {\bibfnamefont {J.}~\bibnamefont {McGuire}},\ }\bibfield  {title} {\bibinfo {title} {\emph {Interacting fermions in one dimension. II. Attractive potential}},\ }\href@noop {} {\bibfield  {journal} {\bibinfo  {journal} {Journal of Mathematical Physics}\ }\textbf {\bibinfo {volume} {7}},\ \bibinfo {pages} {123} (\bibinfo {year} {1966})}\BibitemShut {NoStop}%
\bibitem [{\citenamefont {Guan}\ \emph {et~al.}(2013)\citenamefont {Guan}, \citenamefont {Batchelor},\ and\ \citenamefont {Lee}}]{guan2013fermi}%
  \BibitemOpen
  \bibfield  {author} {\bibinfo {author} {\bibfnamefont {X.-W.}\ \bibnamefont {Guan}}, \bibinfo {author} {\bibfnamefont {M.~T.}\ \bibnamefont {Batchelor}}, \ and\ \bibinfo {author} {\bibfnamefont {C.}~\bibnamefont {Lee}},\ }\bibfield  {title} {\bibinfo {title} {\emph {Fermi gases in one dimension: From Bethe ansatz to experiments}},\ }\href@noop {} {\bibfield  {journal} {\bibinfo  {journal} {Reviews of Modern Physics}\ }\textbf {\bibinfo {volume} {85}},\ \bibinfo {pages} {1633} (\bibinfo {year} {2013})}\BibitemShut {NoStop}%
\bibitem [{\citenamefont {Mahan}(2013)}]{mahan2013many}%
  \BibitemOpen
  \bibfield  {author} {\bibinfo {author} {\bibfnamefont {G.~D.}\ \bibnamefont {Mahan}},\ }\href@noop {} {\emph {\bibinfo {title} {Many-particle physics}}}\ (\bibinfo  {publisher} {Springer Science \& Business Media},\ \bibinfo {year} {2013})\BibitemShut {NoStop}%
\bibitem [{\citenamefont {Combescot}\ \emph {et~al.}(2007)\citenamefont {Combescot}, \citenamefont {Recati}, \citenamefont {Lobo},\ and\ \citenamefont {Chevy}}]{combescot2007normal}%
  \BibitemOpen
  \bibfield  {author} {\bibinfo {author} {\bibfnamefont {R.}~\bibnamefont {Combescot}}, \bibinfo {author} {\bibfnamefont {A.}~\bibnamefont {Recati}}, \bibinfo {author} {\bibfnamefont {C.}~\bibnamefont {Lobo}}, \ and\ \bibinfo {author} {\bibfnamefont {F.}~\bibnamefont {Chevy}},\ }\bibfield  {title} {\bibinfo {title} {\emph {Normal state of highly polarized Fermi gases: simple many-body approaches}},\ }\href@noop {} {\bibfield  {journal} {\bibinfo  {journal} {Physical review letters}\ }\textbf {\bibinfo {volume} {98}},\ \bibinfo {pages} {180402} (\bibinfo {year} {2007})}\BibitemShut {NoStop}%
\bibitem [{\citenamefont {Giraud}\ and\ \citenamefont {Combescot}(2009)}]{giraud2009highly}%
  \BibitemOpen
  \bibfield  {author} {\bibinfo {author} {\bibfnamefont {S.}~\bibnamefont {Giraud}}\ and\ \bibinfo {author} {\bibfnamefont {R.}~\bibnamefont {Combescot}},\ }\bibfield  {title} {\bibinfo {title} {\emph {Highly polarized Fermi gases: One-dimensional case}},\ }\href@noop {} {\bibfield  {journal} {\bibinfo  {journal} {Physical Review A—Atomic, Molecular, and Optical Physics}\ }\textbf {\bibinfo {volume} {79}},\ \bibinfo {pages} {043615} (\bibinfo {year} {2009})}\BibitemShut {NoStop}%
\bibitem [{\citenamefont {Busch}\ \emph {et~al.}(1998)\citenamefont {Busch}, \citenamefont {Englert}, \citenamefont {Rza{\.z}ewski},\ and\ \citenamefont {Wilkens}}]{busch1998two}%
  \BibitemOpen
  \bibfield  {author} {\bibinfo {author} {\bibfnamefont {T.}~\bibnamefont {Busch}}, \bibinfo {author} {\bibfnamefont {B.-G.}\ \bibnamefont {Englert}}, \bibinfo {author} {\bibfnamefont {K.}~\bibnamefont {Rza{\.z}ewski}}, \ and\ \bibinfo {author} {\bibfnamefont {M.}~\bibnamefont {Wilkens}},\ }\bibfield  {title} {\bibinfo {title} {\emph {Two cold atoms in a harmonic trap}},\ }\href@noop {} {\bibfield  {journal} {\bibinfo  {journal} {Foundations of Physics}\ }\textbf {\bibinfo {volume} {28}},\ \bibinfo {pages} {549} (\bibinfo {year} {1998})}\BibitemShut {NoStop}%
\bibitem [{\citenamefont {Ko{\'s}cik}\ and\ \citenamefont {Sowi{\'n}ski}(2018)}]{koscik2018exactly}%
  \BibitemOpen
  \bibfield  {author} {\bibinfo {author} {\bibfnamefont {P.}~\bibnamefont {Ko{\'s}cik}}\ and\ \bibinfo {author} {\bibfnamefont {T.}~\bibnamefont {Sowi{\'n}ski}},\ }\bibfield  {title} {\bibinfo {title} {\emph {Exactly solvable model of two trapped quantum particles interacting via finite-range soft-core interactions}},\ }\href@noop {} {\bibfield  {journal} {\bibinfo  {journal} {Scientific Reports}\ }\textbf {\bibinfo {volume} {8}},\ \bibinfo {pages} {48} (\bibinfo {year} {2018})}\BibitemShut {NoStop}%
\bibitem [{\citenamefont {Ko{\'s}cik}\ and\ \citenamefont {Sowi{\'n}ski}(2019)}]{koscik2019exactly}%
  \BibitemOpen
  \bibfield  {author} {\bibinfo {author} {\bibfnamefont {P.}~\bibnamefont {Ko{\'s}cik}}\ and\ \bibinfo {author} {\bibfnamefont {T.}~\bibnamefont {Sowi{\'n}ski}},\ }\bibfield  {title} {\bibinfo {title} {\emph {Exactly solvable model of two interacting Rydberg-dressed atoms confined in a two-dimensional harmonic trap}},\ }\href@noop {} {\bibfield  {journal} {\bibinfo  {journal} {Scientific Reports}\ }\textbf {\bibinfo {volume} {9}},\ \bibinfo {pages} {12018} (\bibinfo {year} {2019})}\BibitemShut {NoStop}%
\bibitem [{\citenamefont {Sous}\ \emph {et~al.}(2020)\citenamefont {Sous}, \citenamefont {Sadeghpour}, \citenamefont {Killian}, \citenamefont {Demler},\ and\ \citenamefont {Schmidt}}]{sous2020rydberg}%
  \BibitemOpen
  \bibfield  {author} {\bibinfo {author} {\bibfnamefont {J.}~\bibnamefont {Sous}}, \bibinfo {author} {\bibfnamefont {H.}~\bibnamefont {Sadeghpour}}, \bibinfo {author} {\bibfnamefont {T.}~\bibnamefont {Killian}}, \bibinfo {author} {\bibfnamefont {E.}~\bibnamefont {Demler}}, \ and\ \bibinfo {author} {\bibfnamefont {R.}~\bibnamefont {Schmidt}},\ }\bibfield  {title} {\bibinfo {title} {\emph {Rydberg impurity in a Fermi gas: Quantum statistics and rotational blockade}},\ }\href@noop {} {\bibfield  {journal} {\bibinfo  {journal} {Physical Review Research}\ }\textbf {\bibinfo {volume} {2}},\ \bibinfo {pages} {023021} (\bibinfo {year} {2020})}\BibitemShut {NoStop}%
\bibitem [{\citenamefont {Giorgini}\ \emph {et~al.}(2008)\citenamefont {Giorgini}, \citenamefont {Pitaevskii},\ and\ \citenamefont {Stringari}}]{giorgini2008theory}%
  \BibitemOpen
  \bibfield  {author} {\bibinfo {author} {\bibfnamefont {S.}~\bibnamefont {Giorgini}}, \bibinfo {author} {\bibfnamefont {L.~P.}\ \bibnamefont {Pitaevskii}}, \ and\ \bibinfo {author} {\bibfnamefont {S.}~\bibnamefont {Stringari}},\ }\bibfield  {title} {\bibinfo {title} {\emph {Theory of ultracold atomic Fermi gases}},\ }\href@noop {} {\bibfield  {journal} {\bibinfo  {journal} {Reviews of Modern Physics}\ }\textbf {\bibinfo {volume} {80}},\ \bibinfo {pages} {1215} (\bibinfo {year} {2008})}\BibitemShut {NoStop}%
\bibitem [{\citenamefont {Hood}\ \emph {et~al.}(2020)\citenamefont {Hood}, \citenamefont {Yu}, \citenamefont {Lin}, \citenamefont {Zhang}, \citenamefont {Wang}, \citenamefont {Liu}, \citenamefont {Gao},\ and\ \citenamefont {Ni}}]{PhysRevResearch.2.023108}%
  \BibitemOpen
  \bibfield  {author} {\bibinfo {author} {\bibfnamefont {J.~D.}\ \bibnamefont {Hood}}, \bibinfo {author} {\bibfnamefont {Y.}~\bibnamefont {Yu}}, \bibinfo {author} {\bibfnamefont {Y.-W.}\ \bibnamefont {Lin}}, \bibinfo {author} {\bibfnamefont {J.~T.}\ \bibnamefont {Zhang}}, \bibinfo {author} {\bibfnamefont {K.}~\bibnamefont {Wang}}, \bibinfo {author} {\bibfnamefont {L.~R.}\ \bibnamefont {Liu}}, \bibinfo {author} {\bibfnamefont {B.}~\bibnamefont {Gao}}, \ and\ \bibinfo {author} {\bibfnamefont {K.-K.}\ \bibnamefont {Ni}},\ }\bibfield  {title} {\bibinfo {title} {\emph {Multichannel interactions of two atoms in an optical tweezer}},\ }\href {\doibase 10.1103/PhysRevResearch.2.023108} {\bibfield  {journal} {\bibinfo  {journal} {Phys. Rev. Res.}\ }\textbf {\bibinfo {volume} {2}},\ \bibinfo {pages} {023108} (\bibinfo {year} {2020})}\BibitemShut {NoStop}%
\bibitem [{\citenamefont {Backert}\ \emph {et~al.}(2025)\citenamefont {Backert}, \citenamefont {Brauneis}, \citenamefont {\ifmmode~\check{C}\else \v{C}\fi{}ufar}, \citenamefont {Brand}, \citenamefont {Hammer},\ and\ \citenamefont {Volosniev}}]{8mnc-x42q}%
  \BibitemOpen
  \bibfield  {author} {\bibinfo {author} {\bibfnamefont {T.~G.}\ \bibnamefont {Backert}}, \bibinfo {author} {\bibfnamefont {F.}~\bibnamefont {Brauneis}}, \bibinfo {author} {\bibfnamefont {M.}~\bibnamefont {\ifmmode~\check{C}\else \v{C}\fi{}ufar}}, \bibinfo {author} {\bibfnamefont {J.}~\bibnamefont {Brand}}, \bibinfo {author} {\bibfnamefont {H.-W.}\ \bibnamefont {Hammer}}, \ and\ \bibinfo {author} {\bibfnamefont {A.~G.}\ \bibnamefont {Volosniev}},\ }\bibfield  {title} {\bibinfo {title} {\emph {Effective Theory for Strongly Attractive One-Dimensional Fermions}},\ }\href {\doibase 10.1103/8mnc-x42q} {\bibfield  {journal} {\bibinfo  {journal} {Phys. Rev. Lett.}\ }\textbf {\bibinfo {volume} {135}},\ \bibinfo {pages} {040401} (\bibinfo {year} {2025})}\BibitemShut {NoStop}%
\bibitem [{\citenamefont {Silber}\ \emph {et~al.}(2005)\citenamefont {Silber}, \citenamefont {G\"unther}, \citenamefont {Marzok}, \citenamefont {Deh}, \citenamefont {Courteille},\ and\ \citenamefont {Zimmermann}}]{PhysRevLett.95.170408}%
  \BibitemOpen
  \bibfield  {author} {\bibinfo {author} {\bibfnamefont {C.}~\bibnamefont {Silber}}, \bibinfo {author} {\bibfnamefont {S.}~\bibnamefont {G\"unther}}, \bibinfo {author} {\bibfnamefont {C.}~\bibnamefont {Marzok}}, \bibinfo {author} {\bibfnamefont {B.}~\bibnamefont {Deh}}, \bibinfo {author} {\bibfnamefont {P.~W.}\ \bibnamefont {Courteille}}, \ and\ \bibinfo {author} {\bibfnamefont {C.}~\bibnamefont {Zimmermann}},\ }\bibfield  {title} {\bibinfo {title} {\emph {Quantum-Degenerate Mixture of Fermionic Lithium and Bosonic Rubidium Gases}},\ }\href {\doibase 10.1103/PhysRevLett.95.170408} {\bibfield  {journal} {\bibinfo  {journal} {Phys. Rev. Lett.}\ }\textbf {\bibinfo {volume} {95}},\ \bibinfo {pages} {170408} (\bibinfo {year} {2005})}\BibitemShut {NoStop}%
\bibitem [{\citenamefont {Hansen}\ \emph {et~al.}(2011)\citenamefont {Hansen}, \citenamefont {Khramov}, \citenamefont {Dowd}, \citenamefont {Jamison}, \citenamefont {Ivanov},\ and\ \citenamefont {Gupta}}]{PhysRevA.84.011606}%
  \BibitemOpen
  \bibfield  {author} {\bibinfo {author} {\bibfnamefont {A.~H.}\ \bibnamefont {Hansen}}, \bibinfo {author} {\bibfnamefont {A.}~\bibnamefont {Khramov}}, \bibinfo {author} {\bibfnamefont {W.~H.}\ \bibnamefont {Dowd}}, \bibinfo {author} {\bibfnamefont {A.~O.}\ \bibnamefont {Jamison}}, \bibinfo {author} {\bibfnamefont {V.~V.}\ \bibnamefont {Ivanov}}, \ and\ \bibinfo {author} {\bibfnamefont {S.}~\bibnamefont {Gupta}},\ }\bibfield  {title} {\bibinfo {title} {\emph {Quantum degenerate mixture of ytterbium and lithium atoms}},\ }\href {\doibase 10.1103/PhysRevA.84.011606} {\bibfield  {journal} {\bibinfo  {journal} {Phys. Rev. A}\ }\textbf {\bibinfo {volume} {84}},\ \bibinfo {pages} {011606} (\bibinfo {year} {2011})}\BibitemShut {NoStop}%
\bibitem [{\citenamefont {Hara}\ \emph {et~al.}(2011)\citenamefont {Hara}, \citenamefont {Takasu}, \citenamefont {Yamaoka}, \citenamefont {Doyle},\ and\ \citenamefont {Takahashi}}]{PhysRevLett.106.205304}%
  \BibitemOpen
  \bibfield  {author} {\bibinfo {author} {\bibfnamefont {H.}~\bibnamefont {Hara}}, \bibinfo {author} {\bibfnamefont {Y.}~\bibnamefont {Takasu}}, \bibinfo {author} {\bibfnamefont {Y.}~\bibnamefont {Yamaoka}}, \bibinfo {author} {\bibfnamefont {J.~M.}\ \bibnamefont {Doyle}}, \ and\ \bibinfo {author} {\bibfnamefont {Y.}~\bibnamefont {Takahashi}},\ }\bibfield  {title} {\bibinfo {title} {\emph {Quantum Degenerate Mixtures of Alkali and Alkaline-Earth-Like Atoms}},\ }\href {\doibase 10.1103/PhysRevLett.106.205304} {\bibfield  {journal} {\bibinfo  {journal} {Phys. Rev. Lett.}\ }\textbf {\bibinfo {volume} {106}},\ \bibinfo {pages} {205304} (\bibinfo {year} {2011})}\BibitemShut {NoStop}%
\bibitem [{\citenamefont {DeSalvo}\ \emph {et~al.}(2017)\citenamefont {DeSalvo}, \citenamefont {Patel}, \citenamefont {Johansen},\ and\ \citenamefont {Chin}}]{PhysRevLett.119.233401}%
  \BibitemOpen
  \bibfield  {author} {\bibinfo {author} {\bibfnamefont {B.~J.}\ \bibnamefont {DeSalvo}}, \bibinfo {author} {\bibfnamefont {K.}~\bibnamefont {Patel}}, \bibinfo {author} {\bibfnamefont {J.}~\bibnamefont {Johansen}}, \ and\ \bibinfo {author} {\bibfnamefont {C.}~\bibnamefont {Chin}},\ }\bibfield  {title} {\bibinfo {title} {\emph {Observation of a Degenerate Fermi Gas Trapped by a Bose-Einstein Condensate}},\ }\href {\doibase 10.1103/PhysRevLett.119.233401} {\bibfield  {journal} {\bibinfo  {journal} {Phys. Rev. Lett.}\ }\textbf {\bibinfo {volume} {119}},\ \bibinfo {pages} {233401} (\bibinfo {year} {2017})}\BibitemShut {NoStop}%
\bibitem [{\citenamefont {Singh}\ \emph {et~al.}(2022)\citenamefont {Singh}, \citenamefont {Anand}, \citenamefont {Pocklington}, \citenamefont {Kemp},\ and\ \citenamefont {Bernien}}]{PhysRevX.12.011040}%
  \BibitemOpen
  \bibfield  {author} {\bibinfo {author} {\bibfnamefont {K.}~\bibnamefont {Singh}}, \bibinfo {author} {\bibfnamefont {S.}~\bibnamefont {Anand}}, \bibinfo {author} {\bibfnamefont {A.}~\bibnamefont {Pocklington}}, \bibinfo {author} {\bibfnamefont {J.~T.}\ \bibnamefont {Kemp}}, \ and\ \bibinfo {author} {\bibfnamefont {H.}~\bibnamefont {Bernien}},\ }\bibfield  {title} {\bibinfo {title} {\emph {Dual-Element, Two-Dimensional Atom Array with Continuous-Mode Operation}},\ }\href {\doibase 10.1103/PhysRevX.12.011040} {\bibfield  {journal} {\bibinfo  {journal} {Phys. Rev. X}\ }\textbf {\bibinfo {volume} {12}},\ \bibinfo {pages} {011040} (\bibinfo {year} {2022})}\BibitemShut {NoStop}%
\bibitem [{\citenamefont {Anand}\ \emph {et~al.}(2024)\citenamefont {Anand}, \citenamefont {Bradley}, \citenamefont {White}, \citenamefont {Ramesh}, \citenamefont {Singh},\ and\ \citenamefont {Bernien}}]{anand2024dual}%
  \BibitemOpen
  \bibfield  {author} {\bibinfo {author} {\bibfnamefont {S.}~\bibnamefont {Anand}}, \bibinfo {author} {\bibfnamefont {C.~E.}\ \bibnamefont {Bradley}}, \bibinfo {author} {\bibfnamefont {R.}~\bibnamefont {White}}, \bibinfo {author} {\bibfnamefont {V.}~\bibnamefont {Ramesh}}, \bibinfo {author} {\bibfnamefont {K.}~\bibnamefont {Singh}}, \ and\ \bibinfo {author} {\bibfnamefont {H.}~\bibnamefont {Bernien}},\ }\bibfield  {title} {\bibinfo {title} {\emph {A dual-species Rydberg array}},\ }\href@noop {} {\bibfield  {journal} {\bibinfo  {journal} {Nature Physics}\ }\textbf {\bibinfo {volume} {20}},\ \bibinfo {pages} {1744} (\bibinfo {year} {2024})}\BibitemShut {NoStop}%
\bibitem [{\citenamefont {Wei}\ \emph {et~al.}(2024)\citenamefont {Wei}, \citenamefont {Wei}, \citenamefont {Li},\ and\ \citenamefont {Yan}}]{PhysRevA.110.043118}%
  \BibitemOpen
  \bibfield  {author} {\bibinfo {author} {\bibfnamefont {Y.}~\bibnamefont {Wei}}, \bibinfo {author} {\bibfnamefont {K.}~\bibnamefont {Wei}}, \bibinfo {author} {\bibfnamefont {S.}~\bibnamefont {Li}}, \ and\ \bibinfo {author} {\bibfnamefont {B.}~\bibnamefont {Yan}},\ }\bibfield  {title} {\bibinfo {title} {\emph {Dual-species optical tweezer for Rb and K atoms}},\ }\href {\doibase 10.1103/PhysRevA.110.043118} {\bibfield  {journal} {\bibinfo  {journal} {Phys. Rev. A}\ }\textbf {\bibinfo {volume} {110}},\ \bibinfo {pages} {043118} (\bibinfo {year} {2024})}\BibitemShut {NoStop}%
\bibitem [{\citenamefont {Bera}\ \emph {et~al.}(2008)\citenamefont {Bera}, \citenamefont {Bhattacharyya},\ and\ \citenamefont {Bhattacharjee}}]{bera2008perturbative}%
  \BibitemOpen
  \bibfield  {author} {\bibinfo {author} {\bibfnamefont {N.}~\bibnamefont {Bera}}, \bibinfo {author} {\bibfnamefont {K.}~\bibnamefont {Bhattacharyya}}, \ and\ \bibinfo {author} {\bibfnamefont {J.~K.}\ \bibnamefont {Bhattacharjee}},\ }\bibfield  {title} {\bibinfo {title} {\emph {Perturbative and nonperturbative studies with the delta function potential}},\ }\href@noop {} {\bibfield  {journal} {\bibinfo  {journal} {American Journal of Physics}\ }\textbf {\bibinfo {volume} {76}},\ \bibinfo {pages} {250} (\bibinfo {year} {2008})}\BibitemShut {NoStop}%
\bibitem [{\citenamefont {Viana-Gomes}\ and\ \citenamefont {Peres}(2011)}]{viana2011solution}%
  \BibitemOpen
  \bibfield  {author} {\bibinfo {author} {\bibfnamefont {J.}~\bibnamefont {Viana-Gomes}}\ and\ \bibinfo {author} {\bibfnamefont {N.}~\bibnamefont {Peres}},\ }\bibfield  {title} {\bibinfo {title} {\emph {Solution of the quantum harmonic oscillator plus a delta-function potential at the origin: the oddness of its even-parity solutions}},\ }\href@noop {} {\bibfield  {journal} {\bibinfo  {journal} {European journal of physics}\ }\textbf {\bibinfo {volume} {32}},\ \bibinfo {pages} {1377} (\bibinfo {year} {2011})}\BibitemShut {NoStop}%
\bibitem [{\citenamefont {Aky{\"u}z}\ \emph {et~al.}(2024)\citenamefont {Aky{\"u}z}, \citenamefont {Erman},\ and\ \citenamefont {Uncu}}]{akyuz2024harmonic}%
  \BibitemOpen
  \bibfield  {author} {\bibinfo {author} {\bibfnamefont {C.}~\bibnamefont {Aky{\"u}z}}, \bibinfo {author} {\bibfnamefont {F.}~\bibnamefont {Erman}}, \ and\ \bibinfo {author} {\bibfnamefont {H.}~\bibnamefont {Uncu}},\ }\bibfield  {title} {\bibinfo {title} {\emph {The harmonic oscillator potential perturbed by a combination of linear and non-linear Dirac delta interactions with application to Bose--Einstein condensation}},\ }\href@noop {} {\bibfield  {journal} {\bibinfo  {journal} {Physica A: Statistical Mechanics and its Applications}\ }\textbf {\bibinfo {volume} {641}},\ \bibinfo {pages} {129728} (\bibinfo {year} {2024})}\BibitemShut {NoStop}%
\bibitem [{\citenamefont {Donelli}\ \emph {et~al.}(2025)\citenamefont {Donelli}, \citenamefont {Chiara}, \citenamefont {Scazza},\ and\ \citenamefont {Gherardini}}]{donelli2025impactquantumcoherencedynamics}%
  \BibitemOpen
  \bibfield  {author} {\bibinfo {author} {\bibfnamefont {B.}~\bibnamefont {Donelli}}, \bibinfo {author} {\bibfnamefont {G.~D.}\ \bibnamefont {Chiara}}, \bibinfo {author} {\bibfnamefont {F.}~\bibnamefont {Scazza}}, \ and\ \bibinfo {author} {\bibfnamefont {S.}~\bibnamefont {Gherardini}},\ }\href {https://arxiv.org/abs/2508.01444} {\bibinfo {title} {\emph {Impact of quantum coherence on the dynamics and thermodynamics of quenched free fermions coupled to a localized defect}}} (\bibinfo {year} {2025}),\ \Eprint {http://arxiv.org/abs/2508.01444} {arXiv:2508.01444 [cond-mat.quant-gas]} \BibitemShut {NoStop}%
\bibitem [{\citenamefont {Abramowitz}\ and\ \citenamefont {Stegun}(1948)}]{abramowitz1948handbook}%
  \BibitemOpen
  \bibfield  {author} {\bibinfo {author} {\bibfnamefont {M.}~\bibnamefont {Abramowitz}}\ and\ \bibinfo {author} {\bibfnamefont {I.~A.}\ \bibnamefont {Stegun}},\ }\href@noop {} {\emph {\bibinfo {title} {Handbook of mathematical functions with formulas, graphs, and mathematical tables}}},\ Vol.~\bibinfo {volume} {55}\ (\bibinfo  {publisher} {US Government printing office},\ \bibinfo {year} {1948})\BibitemShut {NoStop}%
\bibitem [{Sup()}]{SuppMat}%
  \BibitemOpen
  \href@noop {} {}\bibinfo {note} {See Supplemental Material for the detailed derivation of the main results.}\BibitemShut {Stop}%
\bibitem [{\citenamefont {Fl{\"u}gge}(2012)}]{flugge2012practical}%
  \BibitemOpen
  \bibfield  {author} {\bibinfo {author} {\bibfnamefont {S.}~\bibnamefont {Fl{\"u}gge}},\ }\href@noop {} {\emph {\bibinfo {title} {Practical quantum mechanics}}}\ (\bibinfo  {publisher} {Springer Science \& Business Media},\ \bibinfo {year} {2012})\BibitemShut {NoStop}%
\bibitem [{\citenamefont {Whittaker}\ and\ \citenamefont {Watson}(2020)}]{whittaker2020course}%
  \BibitemOpen
  \bibfield  {author} {\bibinfo {author} {\bibfnamefont {E.~T.}\ \bibnamefont {Whittaker}}\ and\ \bibinfo {author} {\bibfnamefont {G.~N.}\ \bibnamefont {Watson}},\ }\href@noop {} {\emph {\bibinfo {title} {A course of modern analysis}}}\ (\bibinfo  {publisher} {Courier Dover Publications},\ \bibinfo {year} {2020})\BibitemShut {NoStop}%
\bibitem [{\citenamefont {Landau}\ and\ \citenamefont {Lifshitz}(2013)}]{landau2013quantum}%
  \BibitemOpen
  \bibfield  {author} {\bibinfo {author} {\bibfnamefont {L.~D.}\ \bibnamefont {Landau}}\ and\ \bibinfo {author} {\bibfnamefont {E.~M.}\ \bibnamefont {Lifshitz}},\ }\href@noop {} {\emph {\bibinfo {title} {Quantum mechanics: non-relativistic theory}}},\ Vol.~\bibinfo {volume} {3}\ (\bibinfo  {publisher} {Elsevier},\ \bibinfo {year} {2013})\BibitemShut {NoStop}%
\bibitem [{\citenamefont {Barth}\ and\ \citenamefont {Zwerger}(2011)}]{barth2011tan}%
  \BibitemOpen
  \bibfield  {author} {\bibinfo {author} {\bibfnamefont {M.}~\bibnamefont {Barth}}\ and\ \bibinfo {author} {\bibfnamefont {W.}~\bibnamefont {Zwerger}},\ }\bibfield  {title} {\bibinfo {title} {\emph {Tan relations in one dimension}},\ }\href@noop {} {\bibfield  {journal} {\bibinfo  {journal} {Annals of Physics}\ }\textbf {\bibinfo {volume} {326}},\ \bibinfo {pages} {2544} (\bibinfo {year} {2011})}\BibitemShut {NoStop}%
\bibitem [{\citenamefont {Levinsen}\ \emph {et~al.}(2015)\citenamefont {Levinsen}, \citenamefont {Massignan}, \citenamefont {Bruun},\ and\ \citenamefont {Parish}}]{levinsen2015strong}%
  \BibitemOpen
  \bibfield  {author} {\bibinfo {author} {\bibfnamefont {J.}~\bibnamefont {Levinsen}}, \bibinfo {author} {\bibfnamefont {P.}~\bibnamefont {Massignan}}, \bibinfo {author} {\bibfnamefont {G.~M.}\ \bibnamefont {Bruun}}, \ and\ \bibinfo {author} {\bibfnamefont {M.~M.}\ \bibnamefont {Parish}},\ }\bibfield  {title} {\bibinfo {title} {\emph {Strong-coupling ansatz for the one-dimensional Fermi gas in a harmonic potential}},\ }\href@noop {} {\bibfield  {journal} {\bibinfo  {journal} {Science Advances}\ }\textbf {\bibinfo {volume} {1}},\ \bibinfo {pages} {e1500197} (\bibinfo {year} {2015})}\BibitemShut {NoStop}%
\bibitem [{\citenamefont {Anderson}(1967{\natexlab{a}})}]{PhysRevLett.18.1049}%
  \BibitemOpen
  \bibfield  {author} {\bibinfo {author} {\bibfnamefont {P.~W.}\ \bibnamefont {Anderson}},\ }\bibfield  {title} {\bibinfo {title} {\emph {Infrared Catastrophe in Fermi Gases with Local Scattering Potentials}},\ }\href {\doibase 10.1103/PhysRevLett.18.1049} {\bibfield  {journal} {\bibinfo  {journal} {Phys. Rev. Lett.}\ }\textbf {\bibinfo {volume} {18}},\ \bibinfo {pages} {1049} (\bibinfo {year} {1967}{\natexlab{a}})}\BibitemShut {NoStop}%
\bibitem [{\citenamefont {Anderson}(1967{\natexlab{b}})}]{anderson1967infrared}%
  \BibitemOpen
  \bibfield  {author} {\bibinfo {author} {\bibfnamefont {P.~W.}\ \bibnamefont {Anderson}},\ }\bibfield  {title} {\bibinfo {title} {\emph {Infrared catastrophe in Fermi gases with local scattering potentials}},\ }\href@noop {} {\bibfield  {journal} {\bibinfo  {journal} {Physical Review Letters}\ }\textbf {\bibinfo {volume} {18}},\ \bibinfo {pages} {1049} (\bibinfo {year} {1967}{\natexlab{b}})}\BibitemShut {NoStop}%
\bibitem [{\citenamefont {Knap}\ \emph {et~al.}(2012)\citenamefont {Knap}, \citenamefont {Shashi}, \citenamefont {Nishida}, \citenamefont {Imambekov}, \citenamefont {Abanin},\ and\ \citenamefont {Demler}}]{knap2012time}%
  \BibitemOpen
  \bibfield  {author} {\bibinfo {author} {\bibfnamefont {M.}~\bibnamefont {Knap}}, \bibinfo {author} {\bibfnamefont {A.}~\bibnamefont {Shashi}}, \bibinfo {author} {\bibfnamefont {Y.}~\bibnamefont {Nishida}}, \bibinfo {author} {\bibfnamefont {A.}~\bibnamefont {Imambekov}}, \bibinfo {author} {\bibfnamefont {D.~A.}\ \bibnamefont {Abanin}}, \ and\ \bibinfo {author} {\bibfnamefont {E.}~\bibnamefont {Demler}},\ }\bibfield  {title} {\bibinfo {title} {\emph {Time-dependent impurity in ultracold fermions: Orthogonality catastrophe and beyond}},\ }\href@noop {} {\bibfield  {journal} {\bibinfo  {journal} {Physical Review X}\ }\textbf {\bibinfo {volume} {2}},\ \bibinfo {pages} {041020} (\bibinfo {year} {2012})}\BibitemShut {NoStop}%
\bibitem [{\citenamefont {Orso}\ \emph {et~al.}(2025)\citenamefont {Orso}, \citenamefont {Bari{\v{s}}i{\'c}}, \citenamefont {Gradova}, \citenamefont {Chevy},\ and\ \citenamefont {Van~Houcke}}]{orso2025quasi}%
  \BibitemOpen
  \bibfield  {author} {\bibinfo {author} {\bibfnamefont {G.}~\bibnamefont {Orso}}, \bibinfo {author} {\bibfnamefont {L.}~\bibnamefont {Bari{\v{s}}i{\'c}}}, \bibinfo {author} {\bibfnamefont {E.}~\bibnamefont {Gradova}}, \bibinfo {author} {\bibfnamefont {F.}~\bibnamefont {Chevy}}, \ and\ \bibinfo {author} {\bibfnamefont {K.}~\bibnamefont {Van~Houcke}},\ }\bibfield  {title} {\bibinfo {title} {\emph {Quasi-particle residue and charge of the one-dimensional Fermi polaron}},\ }\href@noop {} {\bibfield  {journal} {\bibinfo  {journal} {arXiv preprint arXiv:2504.17558}\ } (\bibinfo {year} {2025})}\BibitemShut {NoStop}%
\bibitem [{\citenamefont {Fogarty}\ \emph {et~al.}(2020)\citenamefont {Fogarty}, \citenamefont {Deffner}, \citenamefont {Busch},\ and\ \citenamefont {Campbell}}]{PhysRevLett.124.110601}%
  \BibitemOpen
  \bibfield  {author} {\bibinfo {author} {\bibfnamefont {T.}~\bibnamefont {Fogarty}}, \bibinfo {author} {\bibfnamefont {S.}~\bibnamefont {Deffner}}, \bibinfo {author} {\bibfnamefont {T.}~\bibnamefont {Busch}}, \ and\ \bibinfo {author} {\bibfnamefont {S.}~\bibnamefont {Campbell}},\ }\bibfield  {title} {\bibinfo {title} {\emph {Orthogonality Catastrophe as a Consequence of the Quantum Speed Limit}},\ }\href {\doibase 10.1103/PhysRevLett.124.110601} {\bibfield  {journal} {\bibinfo  {journal} {Phys. Rev. Lett.}\ }\textbf {\bibinfo {volume} {124}},\ \bibinfo {pages} {110601} (\bibinfo {year} {2020})}\BibitemShut {NoStop}%
\bibitem [{\citenamefont {Gradshteyn}\ and\ \citenamefont {Ryzhik}(2014)}]{gradshteyn2014table}%
  \BibitemOpen
  \bibfield  {author} {\bibinfo {author} {\bibfnamefont {I.~S.}\ \bibnamefont {Gradshteyn}}\ and\ \bibinfo {author} {\bibfnamefont {I.~M.}\ \bibnamefont {Ryzhik}},\ }\href@noop {} {\emph {\bibinfo {title} {Table of integrals, series, and products}}}\ (\bibinfo  {publisher} {Academic press},\ \bibinfo {year} {2014})\BibitemShut {NoStop}%
\end{thebibliography}%

\onecolumngrid
\begin{center}
\newpage
\textbf{
Supplemental Material:\\[4mm]
\large Heavy Fermi polarons in a one-dimensional harmonic trap \\ }

\vspace{4mm}
Nikolay Yegovtsev,$^1$ \\
\vspace{2mm}
{\em \small
$^1$Department of Physics and Astronomy and IQ Initiative, University of Pittsburgh, Pittsburgh, Pennsylvania 15260, USA
}
\end{center} 

\setcounter{equation}{0}
\setcounter{figure}{0}
\setcounter{table}{0}
\setcounter{section}{0}
\setcounter{page}{1}
\makeatletter
\renewcommand{\theequation}{S.\arabic{equation}}
\renewcommand{\thefigure}{S\arabic{figure}}
\renewcommand{\thetable}{S\arabic{table}}
\renewcommand{\thesection}{S.\arabic{section}}
\renewcommand{\theHequation}{S.\arabic{equation}}
\renewcommand{\theHfigure}{S\arabic{figure}}
\renewcommand{\theHtable}{S\arabic{table}}
\renewcommand{\theHsection}{S.\arabic{section}}

\section{Detailed derivation of Eq.~\rf{eq:spectrum}}
Let us consider a two-body problem in 1d, where we have two particles of mass $m$ and $M$ interacting via $\delta$-function potential and each experiencing individual harmonic force:
\begin{equation}
H = -\frac{1}{2m}\frac{\partial^2}{\partial x_1^2} - \frac{1}{2M}\frac{\partial^2}{\partial x_2^2} + g\delta(x_1-x_2) + \frac{m\omega^2x_1^2}{2}+ \frac{M\omega^2x_2^2}{2}.   
\end{equation}
We can introduce the center of mass and relative coordinates $x = x_1-x_2$ and $X = (mx_1+Mx_2)/(m+M)$ to rewrite the Hamiltonian in the form:
\begin{equation}
H= - \frac{1}{2(m+M)}\frac{\partial ^2}{\partial X^2}+\frac{(m+M)\omega^2X^2}{2} -\frac{1}{2\mu}\frac{\partial^2}{\partial x^2}  + \frac{\mu\omega^2x^2}{2}   + g\delta(x),
\end{equation}
where $\mu = \frac{mM}{m+M}$ is the reduced mass. After this separation of variables, the wavefunction becomes a product of wavefunctions of the center of mass and relative motion $\Psi = U(X)\phi(x)$. The center of mass solution is given by the eigenfunctions of a Harmonic oscillator. The relative motion is given by a general combination of two hypergeometric functions so that we can satisfy both the condition of normalizability and the boundary condition at $x=0$. To understand why we need to use hypergeometric functions in the first place, let us quickly revisit the problem of a simple Harmonic oscillator. This part of the discussion closely follows \cite{flugge2012practical}.
\begin{equation}
-\frac{\hbar^2}{2m}\frac{\partial^2}{\partial x^2}\phi + \frac{m\omega^2x^2}{2}\phi = E\phi.  
\end{equation}
Introducing the parameters $k^2=\frac{2mE}{\hbar^2}$ and $\lambda = \frac{m\omega}{\hbar}$, the Schrodinger equation reads:
$$\frac{d^2\phi}{dx^2} + (k^2-\lambda^2 x^2)\phi=0.$$
Asymptotic behavior of the wavefunction for $x\to \infty$ is $\phi\sim e^{-\frac{1}{2}\lambda x^2}$, so factoring this part out as $\phi(x)=e^{-\frac{1}{2}\lambda x^2}v(x)$, we arrive at the equation:
\begin{equation}
\label{eq:vx}
v''-2\lambda x v' + (k^2-\lambda)v=0.    
\end{equation}
Now let us make a change of variables $y=\lambda x^2$, so that the equation becomes of the hypergeometric type:

\begin{equation}
yv'' + \left(\frac{1}{2}-y\right)v'+\left(\frac{k^2}{4\lambda}-\frac{1}{4}\right)v=0.    
\end{equation}
Setting $$\alpha = \frac{1}{4}-\frac{k^2}{4\lambda} = \frac{1}{4}-\frac{E}{2\hbar \omega},$$
the general solution has the form:
\begin{equation}
\label{eq:HOgensol}
v = A F(\alpha ,\frac{1}{2},y) + B y^{1/2}F(\alpha+\frac{1}{2},\frac{3}{2},y).
\end{equation}
The formal Taylor series expansion is given by:
\begin{equation}
\label{eq:cofluent} 
F(\alpha,\gamma, y) = 1 + \frac{\alpha }{\gamma}\frac{y}{1!} + \frac{\alpha(\alpha+1)}{\gamma(\gamma+1)}\frac{y^2}{2!}+\cdots
\end{equation}
In terms of the variable $x$, the first term in the above expression is even, while the second is odd. The asymptotic form of $F(\alpha, \beta, y)$ for $y\to \infty$ is given by:
\begin{equation}
\label{eq:Fasympt}
F(\alpha, \gamma, x)\sim \frac{\Gamma(\gamma)}{\Gamma(\alpha)}e^yy^{\alpha-\gamma},    \end{equation}
so both terms diverge as $e^yy^{\alpha-1/2}$. To make the solution normalizable we need to set the first argument to some negative integer so that the expression reduces to a polynomial. One option is
$$\alpha=-n,\hspace{5mm} E_{2n} = \hbar \omega\left(2n+\frac{1}{2} \right),$$
which gives
\begin{equation}
\phi_{2n}(x) = A F(-n, \frac{1}{2}, \lambda x^2)e^{-\frac{1}{2}\lambda x^2}.    
\end{equation}
Another option is:
$$\alpha+\frac{1}{2}=-n,\hspace{5mm} E_{2n+1} = \hbar \omega\left(2n+\frac{3}{2} \right),$$
which gives
\begin{equation}
\phi_{2n+1}(x) = Bx F(-n, \frac{3}{2}, \lambda x^2)e^{-\frac{1}{2}\lambda x^2}.    
\end{equation}
We see that in 1D, the wavefunctions belong to either the even or the odd sectors. The normalization constants can be computed using the identity from the Appendix in \cite{landau2013quantum}, and the normalized eigenstates are given by:
\begin{equation}
\label{eq:ho0estates}
\begin{split}
&\phi_{2n}(x) = (-1)^n\left[\left(\frac{\lambda}{\pi}\right)^{1/2}\frac{(2n)!}{2^{2n}(n!)^2}\right]^{1/2} F(-n, \frac{1}{2}, \lambda x^2)e^{-\frac{1}{2}\lambda x^2},\\
&\phi_{2n+1}(x) = (-1)^n\left[\frac{\lambda^{3/2}(2n+1)!}{\sqrt{\pi}2^{2n-1}(n!)^2}\right]^{1/2}x F(-n, \frac{3}{2}, \lambda x^2)e^{-\frac{1}{2}\lambda x^2}.
\end{split}
\end{equation}

Returning to the problem at hand, the Schr\"odinger equation for the center-of-mass motion reads:
\begin{equation}
-\frac{\hbar^2}{2\mu}\frac{\partial^2}{\partial x^2}\varphi  + \frac{\mu\omega^2x^2}{2}\varphi   + g\delta(x)\varphi = E\varphi,   
\end{equation}
and we can proceed with the steps as in the analysis of the simple Harmonic oscillator. From the analog of Eq.~\rf{eq:vx} we get:
\begin{equation}
v'(0+)-v'(0-) = \frac{2\mu g}{\hbar^2}v(0),    
\end{equation}
so that the solution should have a discontinuous derivative. This condition only affects the even eigenstates, so the odd ones remain as in the original problem. The solution should also be normalizable, so that the asymptotics of Eq.~\rf{eq:HOgensol} should cancel each other. The combination of solutions from Eq.~\rf{eq:HOgensol} that achieves that goal is given by Tricomi's function \cite{abramowitz1948handbook}:
\begin{equation}
\label{eq:Tricomi}
U(\alpha, \gamma, z) = \frac{\pi}{\sin(\pi \gamma)}\left(\frac{F(\alpha, \gamma, z)}{\Gamma(\alpha-\gamma+1)\Gamma(\gamma)}-z^{1-\gamma}\frac{F(\alpha-\gamma+1,2-\gamma,z)}{\Gamma(\alpha)\Gamma(2-\gamma)}\right).    
\end{equation}
This implies that the wavefunction should be given by:
\begin{equation}
v(x) = AU(\alpha, \frac{1}{2}, \lambda x^2 ).    
\end{equation}
From the expansion in Eq.~\rf{eq:cofluent} and definition Eq.~\rf{eq:Tricomi}, we get:
\begin{equation}
\begin{split}
&v(0) = A\frac{\sqrt{\pi}}{\Gamma(\alpha+\frac{1}{2})},\\
&v'(0+\epsilon) = -v'(0-\epsilon)= -A\frac{2\sqrt{\pi \lambda}}{\Gamma(\alpha)}.
\end{split}    
\end{equation}
Using the expression $\Gamma(z+1)=z\Gamma(z)$, we can finally get the transcendental equation for the spectrum:
\begin{equation}
2\alpha = -\sqrt{\frac{\mu}{\hbar^3\omega}}g\frac{\Gamma(\alpha+\frac{1}{2})}{\Gamma(\alpha+1)}. 
\end{equation}
Taking the limit $M\to \infty$, we arrive at the equation in the main text Eq.~\rf{eq:spectrum}.

\section{Approximate solutions to Eq.~\rf{eq:spectrum}}
\subsection{Weak coupling regime}
In the absence of the interactions $g=0$, the spectrum should be as of the even states of the harmonic oscillator, so $\alpha = -n$. For $\kappa\ll1$ we can seek the solution in the form $\alpha = -n+\kappa c_1+\kappa^2c_2+\cdots$ to the second order in $\kappa$, where $c_n$ are some numbers. Turns out we need to expand both sides to linear order in $\kappa$ to get both $c_1$ and $c_2$:
\begin{equation}
2(-n+c_1\kappa+\cdots) = \frac{(-1)^n}{c_1 (n-1)! \Gamma
   \left(\frac{1}{2}-n\right)}-\frac{(-1)^n  \left(c_1^2 \psi
   ^{(0)}\left(\frac{1}{2}-n\right)-c_1^2 \psi
   ^{(0)}(n)+c_2\right)}{c_1^2 (n-1)! \Gamma
   \left(\frac{1}{2}-n\right)}\kappa    
\end{equation}
From this, we get:
\begin{equation}
\label{eq:c12wc}
\begin{split}
&c_1 = \frac{(-1)^{n+1}}{2n!\Gamma(-n+\frac{1}{2})} =  -\frac{1}{2\pi}\frac{\Gamma(n+\frac{1}{2})}{\Gamma(n+1)},\\
& c_2 = \frac{1}{4\pi^2}\frac{\Gamma^2(n+\frac{1}{2})}{\Gamma^2(n+1)}\left(\frac{1}{n}+\psi^{(0)}(n)-\psi^{(0)}(-n+\frac{1}{2})\right) = \frac{1}{4\pi^2}\left[\frac{\Gamma(n+\frac{1}{2})}{\Gamma(n+1)}\right]^2\left(\psi^{(0)}(n+1)-\psi^{(0)}(n+\frac{1}{2})\right).
\end{split}    
\end{equation}
Plugging this into the expression for the energy, we get the final result in the main text Eq.~\rf{eq:weakcoupling}
\subsection{Strong coupling expansion}
The analysis of negative half-integer scenario is the same both for attractive and repulsive interactions $\alpha = -n\pm\frac{1}{2}+\delta$, where plus corresponds to the attractive and minus to the repulsive interactions, and $\delta$ is the small deviation that we expand in inverse powers of $\kappa$: $\alpha = -n\pm\frac{1}{2}+\frac{c_1}{\kappa} + \frac{c_2}{\kappa^2}+\cdots$.

For the repulsive case, we have:
\begin{equation}
2\left(-n-\frac{1}{2}+\frac{c_1}{\kappa} +\cdots\right) = -(-1)^nn!\Gamma\left(-n+\frac{1}{2}\right)c_1-\frac{(-1)^n n! \Gamma \left(\frac{1}{2}-n\right) \left(c_1^2 \psi
   ^{(0)}\left(\frac{1}{2}-n\right)-c_1^2 \psi ^{(0)}(n+1)+c_2\right)}{\kappa}
\end{equation}
This gives:
\begin{equation}
\label{eq:c12screp}
\begin{split}
&c_1 = \frac{(-1)^n2(n+\frac{1}{2})}{n!\Gamma(-n+\frac{1}{2})} = \frac{2\Gamma(n+\frac{3}{2})}{\pi \Gamma(n+1)},\\
&c_2 = \left[\frac{2\Gamma(n+\frac{3}{2})}{\pi \Gamma(n+1)}\right]^2\left(\psi^{(0)}(n+1)-\psi^{(0)}\left(-n+\frac{1}{2}\right)-\frac{1}{n+\frac{1}{2}}\right) =  \left[\frac{2\Gamma(n+\frac{3}{2})}{\pi \Gamma(n+1)}\right]^2\left(\psi^{(0)}(n+1)-\psi^{(0)}\left(n+\frac{3}{2}\right)\right).
\end{split}    
\end{equation}
Plugging this into the expression for the energy, we get the final result in the main text Eq.~\rf{eq:strongcouplingr}.

For the attractive case, we have:
\begin{equation}
2\left(-n+\frac{1}{2}+\frac{c_1}{\kappa} +\cdots\right) = c_1 (-1)^n (n-1)! \Gamma \left(\frac{3}{2}-n\right)+\frac{(-1)^n (n-1)!
   \Gamma \left(\frac{3}{2}-n\right) \left(c_1^2 \psi
   ^{(0)}\left(\frac{3}{2}-n\right)-c_1^2 \psi ^{(0)}(n)+c_2\right)}{\kappa}.
\end{equation}
This gives:
\begin{equation}
\begin{split}
&c_1 =  -\frac{(-1)^n2(n-\frac{1}{2})}{(n-1)!\Gamma(-n+\frac{3}{2})}    =  \frac{2\Gamma(n+\frac{1}{2})}{\pi \Gamma(n)}, \\
&c_2 = \left[\frac{2\Gamma(n+\frac{1}{2})}{\pi \Gamma(n)}\right]^2\left(\psi^{(0)}(n)-\psi^{(0)}\left(-n+\frac{3}{2}\right)-\frac{1}{n-\frac{1}{2}}\right) = \left[\frac{2\Gamma(n+\frac{1}{2})}{\pi \Gamma(n)}\right]^2\left(\psi^{(0)}(n)-\psi^{(0)}\left(n+\frac{1}{2}\right)\right).
\end{split}    
\end{equation}
Plugging this into the expression for the energy, we get the final result in the main text Eq.~\rf{eq:strongcouplinga}.

\section{Wavefunction normalization and overlaps}
Since we know the form of the wavefunctions, let us compute the overlaps explicitly. Even states are orthogonal to odd, and odd states are orthonormal since did not get modified by the impurity. For the overlap of even states we will need to use the following result \cite{gradshteyn2014table}:
\begin{equation}
\label{eq:UFoverlap}
\begin{split}
&\int_{0}^{\infty}dt\,e^{-t}t^\rho F(a, c,t)U(a',c',\lambda t) = C\frac{\Gamma(c)\Gamma(\beta)}{\Gamma(\gamma)}\lambda^{\sigma}F(c-a,\beta,\gamma,1-\lambda^{-1}),\\
&\rho = c-1,\hspace{5mm} \sigma = -c,\hspace{5mm} \beta=c-c'+1,\hspace{5mm}\gamma = c-a+a'-c'+1,\hspace{5mm}C = \frac{\Gamma(a'-a)}{\Gamma(a')}\\
&\text{or}\\
&\rho = c+c'-2,\hspace{5mm} \sigma = 1-c-c',\hspace{5mm} \beta = c+c'-1,\hspace{5mm} \gamma = a'-a+c,\hspace{5mm} C = \frac{\Gamma(a'-a-c'+1)}{\Gamma(a'-c'+1)}.
\end{split}    
\end{equation}
Using the above expression together with the definition from Eq.~\rf{eq:Tricomi}, we can first compute the normalization constant $A_n$:
\begin{equation}
\begin{split}
&\frac{1}{A_n^2} = \int_{-\infty}^{\infty}dx\,e^{-\lambda x^2}U(\alpha_n,\frac{1}{2},\lambda x^2)U(\tilde{\alpha}_n,\frac{1}{2},\lambda x^2) = \\
&=\frac{\pi}{\sqrt{\lambda}
}\frac{1}{\alpha_n-\tilde{\alpha}_n}\left(\frac{1}{\Gamma(\alpha_n)\Gamma(\tilde{\alpha}_n+\frac{1}{2})}-\frac{1}{\Gamma(\tilde{\alpha}_n)\Gamma(\alpha_n+\frac{1}{2})}\right) \xrightarrow{\tilde{\alpha}_n\to \alpha_n} \frac{\pi}{\sqrt{\lambda}}\left(\frac{\psi
   ^{(0)}\left(\alpha_n+\frac{1}{2}\right) - \psi ^{(0)}(\alpha_n)}{\Gamma (\alpha_n) \Gamma
   \left(\alpha_n+\frac{1}{2}\right)}\right), 
\end{split}
\end{equation}
where $\psi^{(0)}$ is Digamma function:
$$\psi^{(0)}(x) = \frac{\Gamma'(x)}{\Gamma(x)}.$$
Note that in order to obtain the above result we had to regularize the integral by computing the overlap of two functions with different values of $\alpha_n$, and using the definition of $U$ in terms of $F$ to compute the integrals. While both integrals are formally divergent in the limit $\tilde{\alpha}_n\to\alpha_n$ as can be seen from the expression above, their difference is finite and gives the desired answer:
\begin{equation}
\label{eq:AnS}
A_n = \left[\left(\frac{\lambda^{\frac{1}{2}}}{\pi}\right)\frac{\Gamma(\alpha_n)\Gamma(\alpha_n+\frac{1}{2})}{\psi^{(0)}(\alpha_n+\frac{1}{2})-\psi^{(0)}(\alpha_n)}\right]^{1/2},    
\end{equation}
which is one of our results in the main text Eq.~\rf{eq:An}.  

If one is interested in studying the quench dynamics in the above system, one needs to compute the overlaps between the new states in the presence of the impurity and the original eigenstates of the Harmonic oscillator from Eq.~\rf{eq:ho0estates}. The only nontrivial overlaps are between even states. Using Eq.~\rf{eq:UFoverlap}, Eq.~\rf{eq:ho0estates} and Eq.~\rf{eq:AnS} we can compute the following overlap:
\begin{equation}
\label{eq:hooverlapS}
\begin{split}
\langle \varphi_n|\phi_m\rangle &= (-1)^m\left[\left(\frac{\lambda}{\pi^{3/2}}\right)\frac{\Gamma(\alpha_n)\Gamma(\alpha_n+\frac{1}{2})}{\psi^{(0)}(\alpha_n+\frac{1}{2})-\psi^{(0)}(\alpha_n)}\frac{(2m)!}{2^{2m}(m!)^2}\right]^{1/2}\int_{-\infty}^{\infty}dx\,e^{-\lambda x^2}U(\alpha_n,\frac{1}{2},\lambda x^2)F(-m,\frac{1}{2},\lambda x^2) = \\
& = \frac{(-1)^m}{\pi^{1/4}}\left[\frac{\Gamma(\alpha_n)\Gamma(\alpha_n+\frac{1}{2})}{\left(\psi^{(0)}(\alpha_n+\frac{1}{2})-\psi^{(0)}(\alpha_n)\right)}\frac{(2m)!}{2^{2m}(m!)^2}\right]^{1/2}\frac{1}{\Gamma(\alpha_n)}\frac{1}{\alpha_n+m} = \frac{(-1)^m}{\alpha_n+m}\frac{A_n}{\Gamma(\alpha_n)}\left[\left(\frac{\pi}{\lambda}\right)^{\frac{1}{2}}\frac{(2m)!}{2^{2m}m!^2)}\right]^{\frac{1}{2}}.
\end{split}
\end{equation}

In the strong-coupling limit $\kappa\to\infty$, the wave functions of the even sector approach the form of the wave functions of the odd sector for $x>0$, and the matrix elements in Eq.~\rf{eq:hooverlapS} can be computed by substituting $\alpha_n\to-n-1/2$ (one needs to correctly expand the result around $-n$):
\begin{equation}
\label{eq:kappainfoverlaps}
\lim_{\kappa\to\infty}\langle \varphi_n|\phi_m\rangle =  \frac{(-1)^m}{\pi^{1/4}\left(-n+m-\frac{1}{2}\right)\Gamma(-n-\frac{1}{2})}\left[\frac{(2m)!}{2^{2m}(m!)^2}\frac{(-1)^{n+1}\Gamma(-n-\frac{1}{2})}{n!}\right]^{\frac{1}{2}}   
\end{equation}

\section{Quasiparticle residue}
The overlap of the ground states of the system with an infinitely heavy impurity, with the interactions switched on and off, is given by:
\begin{equation}
\label{eq:mnMatrixel}
\langle \Phi|\Phi_\text{NI}\rangle = \frac{1}{(2N)!}\prod_{i=1}^{2N}\,dx_i \det[\varphi^*_1,\varphi^*_2,\dots,\varphi^*_{2N}]\det[\phi_1, \phi_2,\dots, \phi_{2N}] =   
\begin{vmatrix}
\langle\varphi_1|\phi_1\rangle & \cdots &\langle\varphi_1|\phi_{2N}\rangle \\
\vdots & \ddots &\vdots \\
\langle\varphi_{2N}|\phi_1\rangle & \cdots &\langle\varphi_{2N}|\phi_{2N}\rangle
\end{vmatrix}  
\end{equation}
Since we know all the eigenstates, we can compute the one-dimensional integrals analytically. The overlaps of even and odd states are zero by symmetry. Since all odd states in the presence of the impurity are the same as without it, all such states are automatically orthonormal. It is thus convenient to order basis states so that the first $N$ states are even and the remaining $N$ states are odd. In the first $N$ by $N$ block we will have overlaps of the even states, while the second $N$ by $N$ block is diagonal and is equal to the identity, so the total determinant is equal to the determinant of the even sector.
The individual overlaps $\langle \varphi_i|\phi_j\rangle$ can be computed using the result in Eq.~\rf{eq:hooverlapS} for arbitrary values of $\kappa$, and Eq.~\rf{eq:kappainfoverlaps} for the limiting case $\kappa \to \infty$.

\end{document}